\documentclass{emulateapj}
\usepackage{apjfonts}
\begin{document}

\title{On the Production and Survival of Carbon Fuel for Superbursts 
on Accreting Neutron Stars: Implications for Mass Donor Evolution}
\shorttitle{Carbon Survival for Superbursts}

\author{Randall L.\ Cooper, Banibrata Mukhopadhyay, Danny Steeghs, and 
Ramesh Narayan}
\affil{Harvard-Smithsonian Center for Astrophysics, 60 Garden 
Street, Cambridge, MA 02138}

\email{rcooper@cfa.harvard.edu, bmukhopa@cfa.harvard.edu, 
dsteeghs@cfa.harvard.edu, rnarayan@cfa.harvard.edu}


\begin{abstract}

We have investigated the physical conditions under which accreting
neutron stars in low-mass X-ray binaries can both produce and preserve
sufficient quantities of carbon fuel to trigger superbursts.  Our
theoretical models span the plausible ranges of neutron star thermal
conductivities, core neutrino emission mechanisms, and areal radii, as
well as the CNO abundances in the accreted material.  We find that
neutron stars that accrete hydrogen-rich material with CNO mass
fractions $Z_{\mathrm{CNO}} \lesssim Z_{\mathrm{CNO,\odot}}$ will not
exhibit superbursts under any circumstances.  Neutron stars that
accrete material with CNO mass fractions $\gtrsim 4 Z_{\mathrm{CNO,
\odot}}$ will exhibit superbursts at accretion rates in the observed
range.  On this basis, we suggest that the mass donors of superburst
systems must have enhanced CNO abundances.  The accreted CNO acts only
as a catalyst for hydrogen burning via the hot CNO cycle, and
therefore it is the sum of the three elements' mass fractions, not the
individual mass fractions themselves, that is important.  Systems that
exhibit superbursts are observed to differ from those that do not
exhibit superbursts in the nature of their helium-triggered Type I
X-ray bursts: the bursts have shorter durations and much greater
$\alpha$-values.  Increasing the CNO abundance of the accreted
material in our models reproduces both of these observations, thus
once again suggesting enhanced CNO abundances in the mass donors.
Many compact binary systems have been observed in which the abundances
of the accreting material are distinctly non-solar.  Though abundance
analyses of the systems that exhibit superbursts currently do not
exist, Bowen fluorescence blend profiles of 4U 1636-536 and Ser X-1
suggest that the mass donor stars may indeed have non-solar CNO
metallicities.  More detailed abundance analyses of the accreting
matter in systems that exhibit superbursts are needed to verify our
assertion that the matter is rich in CNO elements.

\end{abstract} 


\section{Introduction}

Superbursts are energetic thermonuclear flashes that occur on the
surfaces of accreting neutron stars in low mass X-ray binaries.
They are thought to be caused by unstable carbon burning deep within
the accreted layer.  Superbursts distinguish themselves from their
hydrogen- and helium-burning Type I X-ray burst counterparts (which we
refer to as ``normal'' bursts) by their $\sim 1000$ times larger burst
energies and $\sim 1000$ times longer recurrence times \citep[for
reviews, see][]{C04,SB03}.  As of this writing, nine superbursts have
been observed in seven sources.  Each of these superbursts has an
integrated photon flux of $\approx 10^{42}$ ergs and occurs in a
system with an accretion rate $\dot{M} \approx 0.1$-$0.25
\dot{M}_{\mathrm{Edd}}$, where $\dot{M}_{\mathrm{Edd}}$ denotes the
Eddington limit \citep{K04}.  Four superburst candidates have been
observed in GX 17+2, which has an accretion rate $\dot{M} \approx 0.8
\dot{M}_{\mathrm{Edd}}$ \citep{intZCC04,intZetalHEAD04}.  Very
recently, \citet{RM05} and \citet{K05} have observed two likely 
superbursts in the systems 4U 1608-522 and 4U 0614+091.  The paucity
of observational data makes the recurrence times of superbursts
difficult to determine, though three superbursts have been observed
within 4.7 years from the system 4U 1636-536 \citep{W01,SM02,Ketal04}.

Theoretical studies of superbursts have been rather successful at
reproducing the general observational characteristics, such as the
energetics, recurrence times, and absence of superbursts in systems
with accretion rates below $\approx 0.1 \dot{M}_{\mathrm{Edd}}$
\citep{CB01,SB02,B04,CN04,CN05,CMintZP05}.  Each of these models
requires a carbon mass fraction $\gtrsim 0.1$ at the base of the
accreted layer to trigger a thermonuclear instability.  However, it is
not understood why such a large amount of carbon should exist deep
within the ocean.  Previous theoretical calculations of both
steady-state hydrogen/helium burning \citep{SBCW99,SBCO03} and
unstable helium ignition \citep{Setal01,SBCO03,KHK04,Wetal04,FBLTW05}
on the surface of an accreting neutron star yield far too little
carbon to ignite a superburst.  Consequently, researchers who model
superbursts must set the value of the carbon mass fraction at the base
of the accreted layer ``by hand,'' with little physical motivation.
This is inadequate, since many superburst characteristics, such as the
range of accretion rates in which superbursts occur
\citep{CB01,CN05,CMintZP05} and the amount of energy released by
neutrinos during a superburst \citep{CB01,SB02,CMintZP05}, are strong
functions of the carbon mass fraction.  Thus, no superburst model can
be fully self-consistent until the process by which the carbon fuel
survives both stable and unstable burning is understood.

As of this writing, each system in which a superburst has been
observed exhibits normal Type I X-ray bursts as well
\citep{K04,intZCC04}.  The normal bursts these systems exhibit differ
remarkably from the normal bursts of systems that accrete at similar
rates and from which no superbursts have been observed.  First, the
average e-folding decay times are smaller than those of normal bursts
in systems that do not exhibit superbursts.  Second, for each system
that has exhibited a superburst, $\alpha \gtrsim 1000$, where $\alpha$
is defined as the ratio of the energy released between normal bursts
to the energy released during a normal burst.  This is significantly
greater than the $\alpha$-values of most systems with similar
accretion rates and in which no superbursts have been detected
\citep{Ketal02,intZetal03,intZetalHEAD04}.  For systems with high
accretion rates and in which the accreted matter is predominantly
hydrogen, \citet{NH03} found that normal bursts occur in a unique
regime that they refer to as ``delayed mixed bursts.''  In these
systems, a large fraction of the hydrogen and helium fuel burns stably
before the full thermal instability is triggered.  This stable burning
explains the observations of \citet{vPPL88} and \citet{Cetal03} who
found that, for systems that accrete at a high rate and exhibit such
Type I X-ray bursts, the quantity $\alpha$ rises dramatically to
values $\gtrsim 1000$.  \citet{CN05} hypothesized that delayed mixed
bursts may be the source of the substantial amount of carbon fuel
needed to trigger a superburst in systems for which the accreted
material is predominantly hydrogen.  However, $\alpha$-values of the
delayed bursts of \citet{NH03} reach $\sim 1000$ only if $\dot{M} \sim
0.3 \dot{M}_{\mathrm{Edd}}$.  For $\dot{M} \sim 0.1
\dot{M}_{\mathrm{Edd}}$, roughly the lowest accretion rate at which
delayed mixed bursts occur, they find $\alpha \lesssim 100$.  The low
$\alpha$-values for $\dot{M} \sim 0.1 \dot{M}_{\mathrm{Edd}}$ are
consistent with the results of previous and subsequent unstable helium
ignition studies, all of which yield a negligible amount of carbon.
As of yet, no model has reproduced the observed characteristics of
normal bursts from systems that exhibit superbursts.

Any successful theoretical model of the low mass X-ray binaries
considered in this paper must reproduce the following four phenomena
that are unique to these systems.  First, it must reproduce
superbursts with energies and recurrence times that are consistent
with observations.  Second, nuclear burning in the oceans must produce
sufficient amounts of carbon, and the carbon must survive until a
superburst ignites.  Third, delayed normal bursts with $\alpha \gtrsim
1000$ must occur.  Fourth, the normal bursts that occur must have
unusually short durations.  Previous authors have made significant
progress on the first phenomenon, but none has successfully addressed
the latter three.  In this investigation, we construct models that
attempt to explain all four phenomena.  We begin in \S\ref{themodel}
with a description of our theoretical model.  In \S\ref{Csurvival} we
discuss the physical conditions under which sufficient amounts of
carbon may survive long enough to ultimately fuel a superburst.  We
discuss the results of our model in \S\ref{results}, and we compare
our results to those of previous theoretical studies in
\S\ref{comparewth}.  In \S\ref{comparewobs} we compare our results
with observations.  Our model predicts that the matter accreted onto
the surface of the neutron star must be rich in CNO in order for
superbursts to occur.  We discuss the observational evidence that such
an overabundance exists in \S\ref{CNOenhancement}, and we conclude
with a summary in \S\ref{summary}.

\section{The Model}\label{themodel}

In this investigation, we assume that matter accretes spherically onto
a neutron star of gravitational mass $M$ and areal radius $R$ at a
rate $\dot{M}$, where $\dot{M}$ is the rest mass accreted per unit
time as measured by an observer at infinity.  We set the hydrogen and
heavy element composition of the accreted matter to be that of the
Sun, such that at the stellar surface the hydrogen mass fraction
$X_{\mathrm{out}} = 0.7$ and the heavy element fraction
$Z_{\mathrm{out}} = 0.004$, where $Z$ refers to all metals other than
CNO.  We set the helium mass fraction $Y_{\mathrm{out}} = 1 -
X_{\mathrm{out}} - Z_{\mathrm{out}} - Z_{\mathrm{CNO, out}}$, where
$Z_{\mathrm{CNO, out}}$ is the mass fraction of CNO elements accreted
from the companion star.  We treat $Z_{\mathrm{CNO, out}}$ as a free
parameter.

\subsection{Thermal and Hydrostatic Structure of the Crust}\label{thermstruc}

To calculate the equilibrium configuration of the neutron star crust,
we use the theoretical model of \citet{CN05} with two modifications.
To calculate the inner temperature boundary condition at the
crust-core interface, the authors assume that the core emits neutrinos
via either modified Urca (mUrca) reactions or pionic reactions.  They
use the neutrino luminosities from \citet{ST83} and thereby determine
the core temperature.  The core temperatures resulting from these
neutrino luminosities are $\sim 3 \times 10^{8}$ K and $\sim 2 \times
10^{7}$ K, respectively.  However, if baryons exist in the core, they
are probably superfluid at high densities \citep{BPP69}.  Baryon
superfluidity greatly suppresses neutrino emission via mUrca reactions
and therefore raises the core temperature \citep{YLS99,YKGH01,YP04}.
Thus, when we model the effects of baryon superfluidity, we set
the neutrino luminosity to be
\begin{equation}
L_{\nu} = (2.0 \times 10^{37} \mathrm{ergs\,s}^{-1})
\frac{m}{M_{\odot}} \left (\frac{T}{10^{9} \mathrm{K}} \right )^{8},
\end{equation}
where $m$ is the interior gravitational mass and $T$ is the proper
temperature, both evaluated at the crust-core interface.  The core
temperatures resulting from this neutrino luminosity are $\sim 6
\times 10^{8}$ K.  Note that we set the coefficient of $L_{\nu}$ not
to correspond to any specific emissivity model, but rather to produce
core temperatures in the desired range.

\citet{CN05} define the free parameter $C_{\mathrm{f}}$ as
the fraction of hydrogen and helium that ultimately burns to carbon.
This parameter enables the authors to set the value of
$Z_{\mathrm{CNO,base}}$, the mass fraction of carbon at the base of
the accreted layer, such that
\begin{equation}
Z_{\mathrm{CNO,base}} = Z_{\mathrm{CNO,out}} +
C_{\mathrm{f}}(X_{\mathrm{out}} + Y_{\mathrm{out}}).
\end{equation}
The value of $C_{\mathrm{f}}$ should affect not only the carbon yield,
but also the amount of energy generated within the crust, because the
total energy released per unit mass when helium is burned ultimately
to iron peak elements is greater than the energy released when helium
is burned only to carbon.  To account for the additional energy
generation, we include $C_{\mathrm{f}}$ in the energy conservation
equation.  Thus, equation (6) of \citet{CN05} becomes
\begin{eqnarray}\label{energyequation}
e^{-2 \Phi/c^{2}} \frac{{\partial}}{{\partial}\Sigma}\left (\frac{F
	r^{2}}{R^{2}} e^{2 \Phi/c^{2}}\right ) = T
	\frac{{\mathrm{d}}s}{{\mathrm{d}}t} - 
\nonumber\\ \nonumber \\
\left (\epsilon_{\mathrm{H}} +
	\left [1 + \frac{E^{*}_{\mathrm{C}}}{E^{*}_{\mathrm{He}}}
	(1-C_{\mathrm{f}}) \right ] \epsilon_{\mathrm{He}} +
	\epsilon_{\mathrm{C}} + \epsilon_{\mathrm{N}} -
	\epsilon_{\mathrm{\nu}} \right ).
\nonumber \\
\end{eqnarray}

\subsection{Nucleosynthesis in the Accreted Layer}

The thermal and hydrostatic structure model described above is quite
sophisticated, approaching the state-of-the-art of such
one-dimensional models.  Its nuclear reaction network, however, is
quite simplistic, since it includes reaction rates only for hydrogen,
helium, and carbon burning.  The hot nuclear flow through stable
thermonuclear burning on the surface of an accreting neutron star
almost certainly produces a wide assortment of different isotopes
through hydrogen and triple-$\alpha$ reactions, rp- and
$\alpha$p-processes, and $\alpha$-captures.  To calculate the detailed
nuclear flow due to stable thermonuclear burning, we consider a large
number of isotopes on either side of the stability line. Our nuclear
reaction network contains $255$ nuclear species up to $^{72}$Ge and
all of the possible reactions between the various isotopes.  Stable
burning at the sub-Eddington accretion rates we consider in this
investigation produces negligible abundances of isotopes with atomic
weights larger than $72$ \citep[e.g.][]{SBCW99}, so the size of our
network is adequate for our purposes.  Lists of the major nuclear
reactions that may take place in the flow are given in previous work
\citep[e.g.][]{C83,L99}.  In our nucleosynthesis code, we use the
reaction rates of \citet{FCZ75} but include the updated rates of
\citet{WFH67}, \citet{FFN80,FFN82a,FFN82b}, \citet{T80}, \citet{WW81},
and \citet{HFCZ83}.  Each of our reaction rates has been updated and
is current up to the year 1996 or later (F.-K.\ Thielemann 1996,
private communication).  The code has been successfully used by one of
us in previous studies of thermonuclear reactions in hot accretion
discs \citep{CM99,MC00,MC01}.  Therefore, we are confident that this
code is applicable to our present investigation.

The nuclear reaction network, which is a set of coupled differential
equations, is linearized and evolved in time along the thermal and
hydrostatic structure of the ocean derived from the model described in
\S\ref{thermstruc}.  This well-proven method is widely used in the
literature \citep[e.g.][]{AT69,WAC73,MC00}.  Here, we briefly outline
how we time-evolve the isotopic abundances.  For simplicity, we
consider only four isotopes and only three reactions:
$^{1}$H($p$,$\beta^{+}\nu$)D, D(D,$\gamma$)${^4}$He, and $2\mathrm{\,}
{^4}$He($\alpha$,$\gamma$)$^{12}$C, although, in our actual
calculations, we use $255$ isotopes and all of the possible reactions
between the isotopes as mentioned above.  Neglecting any backward
reactions, the corresponding rate equations can be expressed as
\begin{eqnarray} \frac{\mathrm{d}}{\mathrm{d} t}\left(
\begin{array}{cccc}X_{\mathrm{H}} \\ X_{\mathrm{D}} \\ X_{\mathrm{He}} \\ X_{\mathrm{C}} \end{array}\right)=
\left(\begin{array}{cccc} -\lambda_{\mathrm{H}} & 0 & 0 & 0\\
\lambda_{\mathrm{H}} & -\lambda_{\mathrm{D}} & 0 & 0 \\ 0 &
\lambda_{\mathrm{D}} & -\lambda_{\mathrm{He}} & 0\\ 0 & 0 &
\lambda_{\mathrm{He}} & 0\end{array}\right)
\left(\begin{array}{cccc}X_{\mathrm{H}} \\ X_{\mathrm{D}} \\
X_{\mathrm{He}} \\ X_{\mathrm{C}} \end{array}\right), \label{nucreac}
\end{eqnarray} 
where the various $\lambda$'s and $X$'s are the reaction rates and
mass fractions of the isotopes, respectively. The above equation can
be written as
\begin{eqnarray}
\frac{\mathrm{d {\bf v}} }{\mathrm{d} t}={\bf \Lambda} \mathrm{{\bf v}}.
\label{nucreac2} 
\end{eqnarray} 
If ${\bf \Lambda}$ is diagnolizable, we can solve this equation by
finding the four eigenvalues $\lambda_{i}$ and four eigenvectors
$\mathrm{{\bf u}}_{i}$ of ${\bf \Lambda}$.  $\mathrm{{\bf v}}(t)$ is
then a linear combination of the eigenvectors such that
\begin{eqnarray}
\mathrm{{\bf v}}(t)=\sum_{i} A_{i} e^{\lambda_{i} t}\mathrm{{\bf u}}_{i},
\label{vt} 
\end{eqnarray} 
where one determines the constant coefficients $A_{i}$ by setting
$\mathrm{{\bf v}}(0) = \mathrm{{\bf v}}_{0}$, the initial elemental
abundance.  Note that this method is correct only if the individual
nuclear lifetimes are constant.
In our actual computational code, we include $255$ isotopes, so ${\bf
\Lambda}$ is a $255\times 255$ matrix with $255$ eigenvalues and $255$
eigenvectors, and thus $i$ runs from $1$ to $255$.

To evolve the composition of the entire layer in time, we perform this
procedure at each timestep.  The reaction rates are in general
functions of the temperature, density, and composition, and therefore
they must be recalculated at each timestep.  Normally, we set the
timestep $\Delta t = \Delta \Sigma / \dot{\Sigma}$, where $\Delta
\Sigma$ is the thickness (mass per unit area) of a thin spherical
shell and $\dot{\Sigma}$ is the mass accretion rate per unit area.
However, we restrict the fractional change in the abundance of each
species $i$ to be less than $\delta$, which we have set to $0.05$ for
our calculations.  Specifically, for each species $i$ such that its
mass fraction $X_{i} \ge 10^{-20}$, we require $\Delta t < \delta\,
{\rm min}[ X_i/(\mathrm{d}X_i/\mathrm{d}t)]$.  Note that the
fractional change in the abundance of each species with a nontrivial
mass fraction is always much less than $0.05$.  Nonetheless, we have
performed calculations for values of $\delta$ both somewhat greater
than and much less than $0.05$, and we have obtained the same results.
Therefore, any error in our results due to our choice of stepsize is
negligible compared to that due to the uncertainties in the reaction
cross sections.  We have verified that our nuclear reaction network
conserves mass to within one part in $10^{8}$.

\subsection{Carbon Abundance in the Accreted Layer}\label{Cabundance}

To calculate the mass fraction of carbon that exists at the base of
the accreted layer, we proceed as follows.  First, we calculate the
equilibrium configuration of the neutron star crust as described in
\S\ref{thermstruc}.  To perform this calculation, we choose an
arbitrary value for the carbon mass fraction at the base of the layer.
This value will be refined later.  Second, we use this equilibrium
configuration in our nucleosynthesis model to calculate the true
carbon mass fraction at the base of the layer.  In general, this value
will be different than that used to determine the equilibrium
configuration.  Therefore, we repeat this two-step process, but we use
the refined carbon mass fraction derived from our nucleosynthesis code
to calculate the equilibrium configuration.  We iterate this process
until successive values of the carbon mass fraction determined from
our nucleosynthesis model agree to sufficient accuracy.  This process
converges rapidly, and the value to which it converges is completely
insensitive to the value of our initial guess.  As an example of a typical
convergence, the carbon mass fraction after each iteration for a
particular calculation, starting with an initial guess of $0.2$, is
$0.52709$, $0.56132$, $0.56477$, $0.56521$, and $0.56531$.  After we
have determined the carbon mass fraction, we conduct the full global
linear stability analysis of \citet{CN05} to determine whether a
superburst will occur, assuming that all of the carbon produced via
stable burning has survived (see \S\ref{Csurvival}).

The reader should note that we do not couple the nuclear energy
generation rates of our reaction network directly into the energy
conservation equation (\ref{energyequation}).  In general, the
contributions from hydrogen and helium burning dominate the total
nuclear energy generation rate, and we calculate these contributions
in equation (\ref{energyequation}) to high accuracy.  Nevertheless, to
ensure the internal consistency of our method, we have performed
calculations in which the energy generation rates of our reaction
network are coupled into equation (\ref{energyequation}), and we find
that the differences in the final carbon mass fractions are
insignificant.

\section{Survival of the Carbon Fuel}\label{Csurvival}

Helium fusion via the triple-$\alpha$ reaction produces most of the
carbon fuel that eventually triggers a superburst.  Uncertainties in
the carbon abundance deep within the accreted layer lie not in carbon's
production, but in its survival.  To ultimately become fuel for a
superburst, carbon must survive both the stable burning
that produces it and the unstable burning during normal Type
I X-ray bursts that potentially consumes it.

\subsection{Survival During Stable Burning}\label{stableburn}

The rapid proton (rp) process of \citet{WW81} is the primary culprit
in the destruction of $^{12}$C.  If hydrogen is present, hydrogen will
burn via the hot CNO cycle \citep{HF65}.  During the hot CNO cycle,
essentially all of the CNO elements will be processed into $^{14}$O
and $^{15}$O.  For the sub-Eddington accretion rates considered in
this paper, the dominant breakout reaction from the hot CNO cycle into
the rp-process is $^{15}$O($\alpha$, $\gamma$)$^{19}$Ne
\citep{WW81,SBCW99,FGWD04}.  It is possible for $^{19}$Ne to return to
the hot CNO cycle by the series of reactions $^{19}$Ne($\beta^{+}
\nu$)$^{19}$F($p$, $\alpha$)$^{16}$O($p$, $\gamma$)$^{17}$F($p$,
$\gamma$)$^{18}$Ne($\beta^{+} \nu$)$^{18}$F($p$, $\alpha$)$^{15}$O,
but if $^{19}$Ne captures a proton by the reaction $^{19}$Ne($p$,
$\gamma$)$^{20}$Na, the ion can never return to the CNO cycle
\citep{WW81}.  Therefore, a $^{12}$C ion that is either accreted from
the companion star or produced from the triple-$\alpha$ reaction
will not survive if it is both converted to $^{15}$O in the hot CNO
cycle and removed from the hot CNO cycle via breakout reactions.

For the relatively high accretion luminosities considered in this
investigation, the rate at which hydrogen burns is set by the
$\beta$-decay timescales of $^{14}$O and $^{15}$O, and therefore the
rate is both temperature- and density-independent.  The hydrogen
nuclear energy generation rate for the hot CNO cycle is
$\epsilon_{\mathrm{H}} = 6 \times 10^{15} Z_{\mathrm{CNO}}$
$\mathrm{ergs\, g}^{-1} \mathrm{\,s}^{-1}$ \citep{HF65}.  The column
depth (mass per unit area) at which hydrogen burns out,
$\Sigma_{\mathrm{H}}$, is thus
\begin{equation}
\Sigma_{\mathrm{H}} \approx \dot{\Sigma} \frac{X_{\mathrm{out}}
E^{*}_{\mathrm{H}}}{\epsilon_{\mathrm{H}}} \approx (\dot{\Sigma}
\times 1100 \mathrm{\,s}) \frac{X_{\mathrm{out}}} {Z_{\mathrm{CNO,
out}}},
\label{sigmaH}
\end{equation}
where $\dot{\Sigma}$ is the mass accretion rate per unit area and
$E^{*}_{\mathrm{H}}$ is the total nuclear energy released per unit
mass of hydrogen burned.  In general, the column depth at which
hydrogen burns increases with increasing accretion rate.

In contrast, the rate at which helium burns is very sensitive to both
temperature and density.  The temperatures of the envelope and ocean
of an accreting neutron star typically increase as the accretion rate
increases.  Consequently, for a pure column of helium, the column
depth at which helium burns decreases with increasing accretion rate.
This statement regarding the depth at which helium burns is not
strictly true for a hydrogen/helium mixture because the hydrogen
burning will steadily add helium to the matter, but the general idea
is still valid.

\begin{figure*}
\plottwo{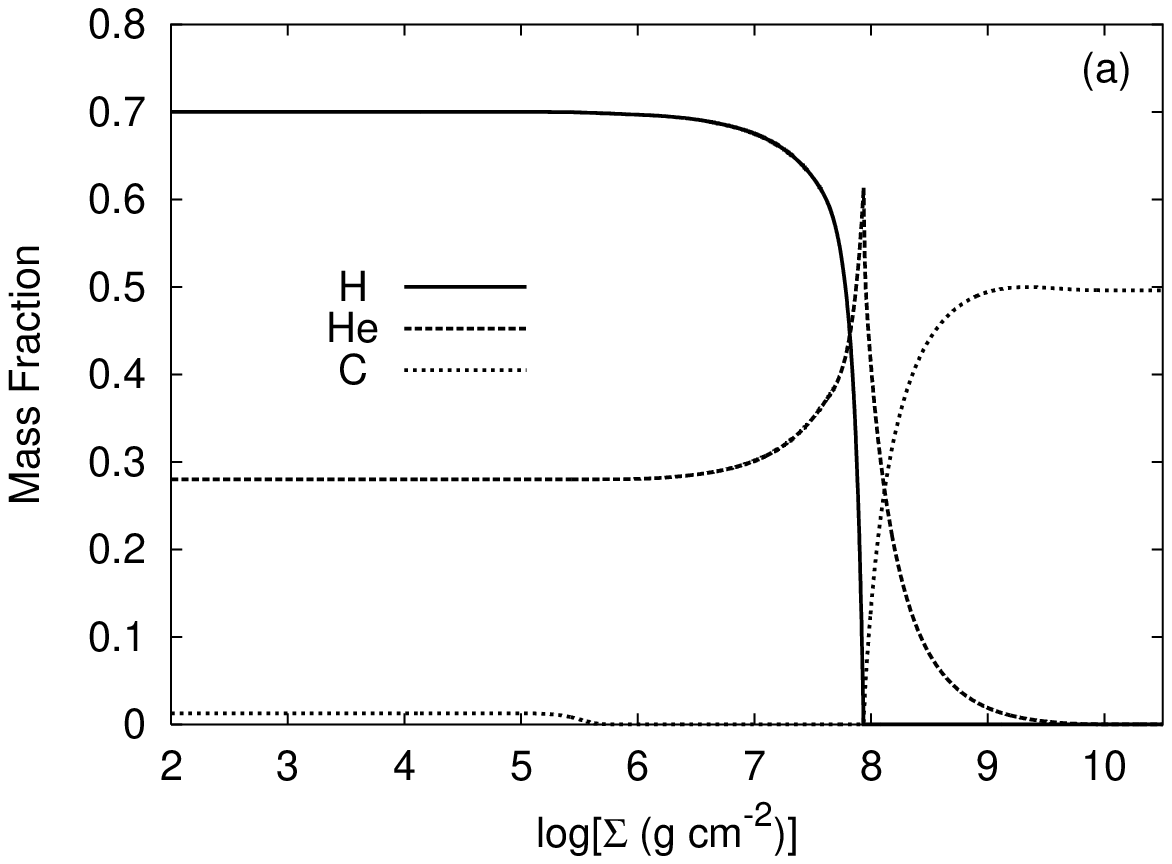}{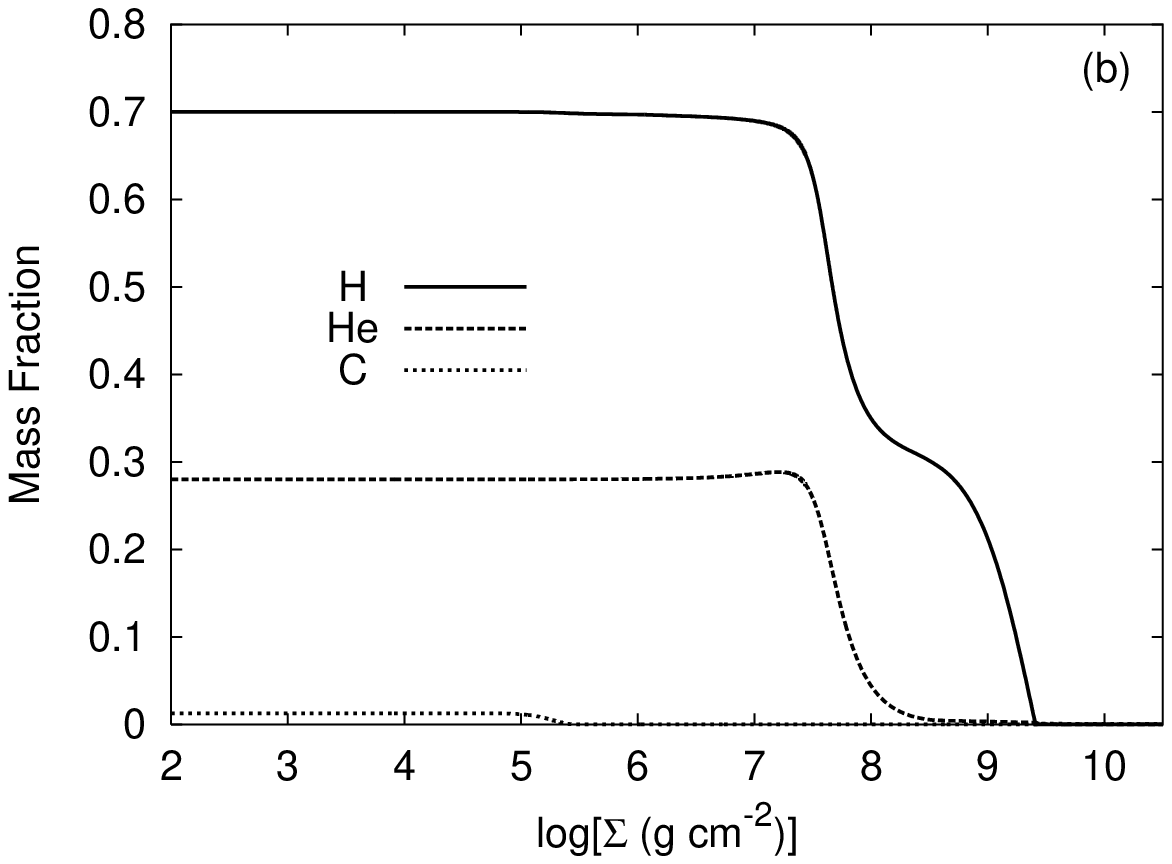}
\caption{$^{1}$H, $^{4}$He, and $^{12}$C mass fractions in the
accreted column due to stable burning for two neutron stars with
different accretion rates.  $\dot{M} = 0.1 \dot{M}_{\mathrm{Edd}}$ for
Figure (a) and $\dot{M} = 0.3 \dot{M}_{\mathrm{Edd}}$ for Figure (b).
Any $^{12}$C accreted from the companion star is depleted when $^{1}$H
begins to burn because $^{12}$C is processed into $^{14}$O and
$^{15}$O during the hot CNO cycle.  For $\dot{M} = 0.1
\dot{M}_{\mathrm{Edd}}$, $^{1}$H burns out before most of the $^{4}$He
has started to burn, so much of the $^{12}$C produced via
triple-$\alpha$ reactions survives.  For $\dot{M} = 0.3
\dot{M}_{\mathrm{Edd}}$, $^{1}$H and $^{4}$He burn simultaneously, so
the $^{12}$C produced via triple-$\alpha$ reactions is lost due to
CNO cycle breakout reactions.
}
\label{HHeCburning}
\end{figure*}

If a large amount of carbon is to survive episodes of stable burning
to eventually trigger a superburst, most of the accreted hydrogen must
burn via the hot CNO cycle before helium starts to burn.  If this
occurs, then the carbon produced via helium burning will not be
processed into $^{15}$O and leave the CNO cycle through rp-process
breakout reactions.  Since the depth at which hydrogen burns increases
with increasing accretion rate, and the depth at which helium burns
decreases with increasing accretion rate, one would expect that more
carbon will survive episodes of stable burning at lower accretion
rates \citep{SBCW99,SBCO03}.  Figure \ref{HHeCburning} illustrates
this trend.

\subsection{Survival During Unstable Burning}

Any carbon that survives stable burning must also withstand unstable
burning in normal Type I bursts if the carbon is to ultimately become
fuel for a superburst.  Previous time-dependent models of normal
prompt bursts, for which $\alpha \lesssim 100$, imply that essentially
no carbon will survive unstable hydrogen and helium burning
\citep{Setal01,SBCO03,KHK04,Wetal04,FBLTW05}.  \citet{CN05} speculate
that delayed bursts, which are bursts that are triggered after a long
period of stable burning has occurred and thus result in extremely
high $\alpha$-values, will leave most of the thick layer of carbon
that exists below the ignition region intact.  Presumably, the
increase in temperature due to hydrogen and helium burning in the
delayed burst is insufficient to ignite most of the carbon that exists
deep in the ocean.  Although we are currently unable to either confirm 
or refute this assertion, we note that observations do imply that 
a sufficient amount of carbon does indeed survive delayed bursts 
\citep{intZetal03}.  \citet{Wetal04} show that the ashes of a given
normal burst are reprocessed in subsequent bursts.  This
``compositional inertia'' may destroy some fraction of the carbon that
a given burst leaves intact.  However, \citet{Wetal04} find that the
effects of compositional inertia are diminished if the CNO abundance
of the accreted material is high or if the burst recurrence time is
long, both of which apply for the models we will suggest later in
\S\ref{results}.  Unfortunately, no detailed time-dependent studies of
the delayed bursts observed in these systems have been carried out as
of this writing.  Therefore, our speculation that a large fraction 
of the carbon produced via stable burning will survive 
delayed bursts must be investigated
further.  For the purposes of this work, we assume that, if carbon is
produced via stable burning, it will be destroyed if a normal prompt
burst ignites (for which $\alpha \lesssim 100$), and it will survive
if either a normal delayed burst ignites (for which $\alpha \gg 100$)
or the system is stable to normal bursts.

\section{Results}\label{results}

We have constructed a total of 144 different models to determine the
conditions under which accreting neutron stars both produce and
preserve sufficient amounts of carbon to trigger superbursts.  We
choose four different stellar areal radii, three different core
neutrino emission mechanisms, and three different conductive opacity
prescriptions, which likely bracket the true radii, core neutrino
emissivities, and conductive opacities of neutron stars found in
nature.  See Table 1 for a list of these parameters.  The column
``Core $\nu$ Emissivity'' describes the neutrino emission mechanism in
the core, where ``hot mUrca'' refers to a stellar core that cools via
mUrca reactions suppressed by baryon superfluidity (see
\S\ref{thermstruc}), ``mUrca''' refers to a core that cools via mUrca
reactions, and ``Pion'' refers to a core that cools via pionic
reactions.  The column ``$Q$'' describes the conductive opacity of the
crust.  For this column, ``$5.2$'' and ``$100$'' refer to neutron
stars with inner crusts that have formed ordered crystal lattices and
which have impurity parameter values \citep{IK93,B00} of $5.2$ and
$100$, respectively, and ``disordered'' refers to a neutron star with
a completely disordered crust.  See \citet{B04} and \citet{CN05} for
details regarding these parameters and their effects upon superbursts
characteristics.  Additionally, we choose four different values of
$Z_{\mathrm{CNO, out}}$, the mass fraction of CNO elements accreted
from the companion star.

A superburst will occur if a large amount of carbon deep within the
ocean undergoes unstable thermonuclear fusion.  In this investigation,
we say that a superburst occurs if the carbon produced by stable
helium burning survives (see \S\ref{stableburn}) and the resulting
carbon-rich accreted column is unstable according to the
general-relativistic global linear stability analysis of \citet{CN05}.
For definiteness, we use the criterion that the carbon yield derived
from our nucleosynthesis model will not survive if a normal burst
occurs with $\alpha < 500$.  This cutoff value is close to that of the
superbursting system with the smallest $\alpha$-value, 4U 1636-536,
for which $\alpha \approx 440$ \citep{intZetalHEAD04}.  The precise
value of this cutoff is unimportant for the final results.

\subsection{Results for $Z_{\mathrm{CNO, out}} = Z_{\mathrm{CNO, \odot}}$}
\label{resultssolar}

\begin{figure*}
\plottwo{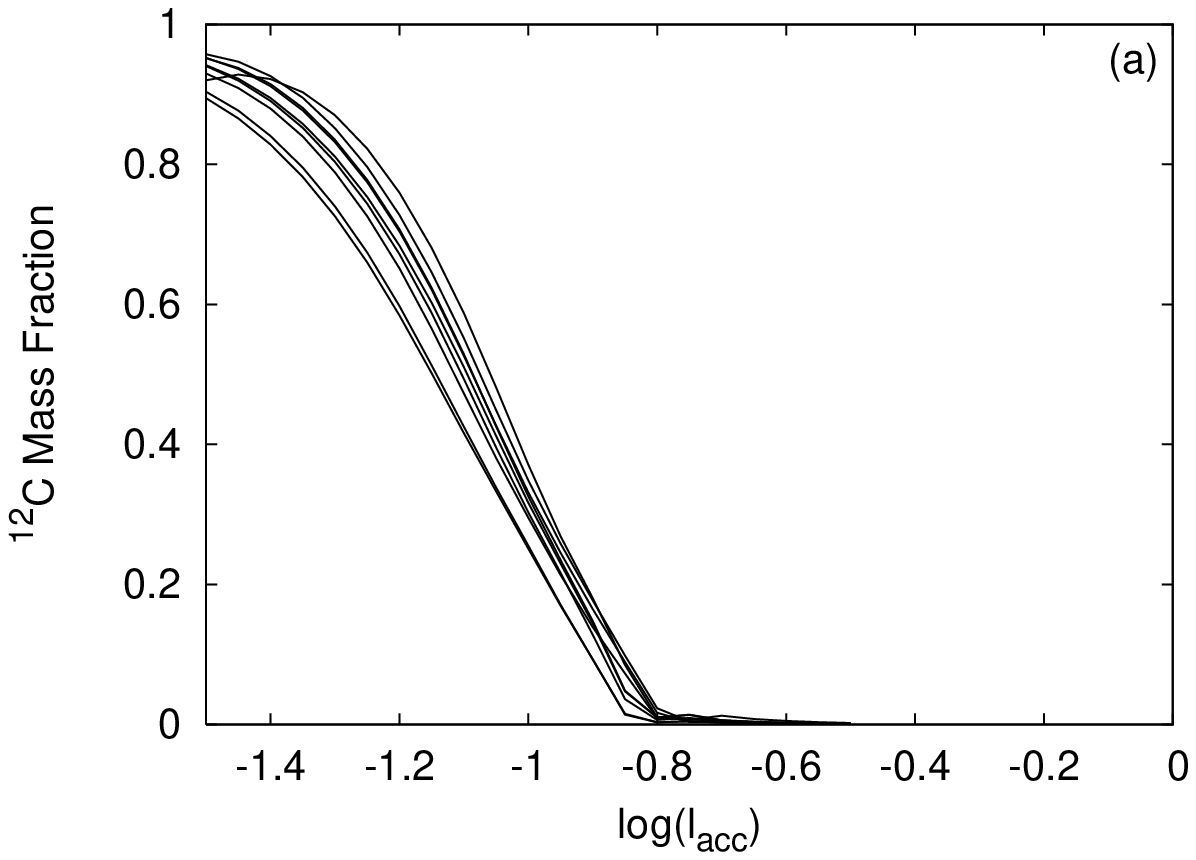}{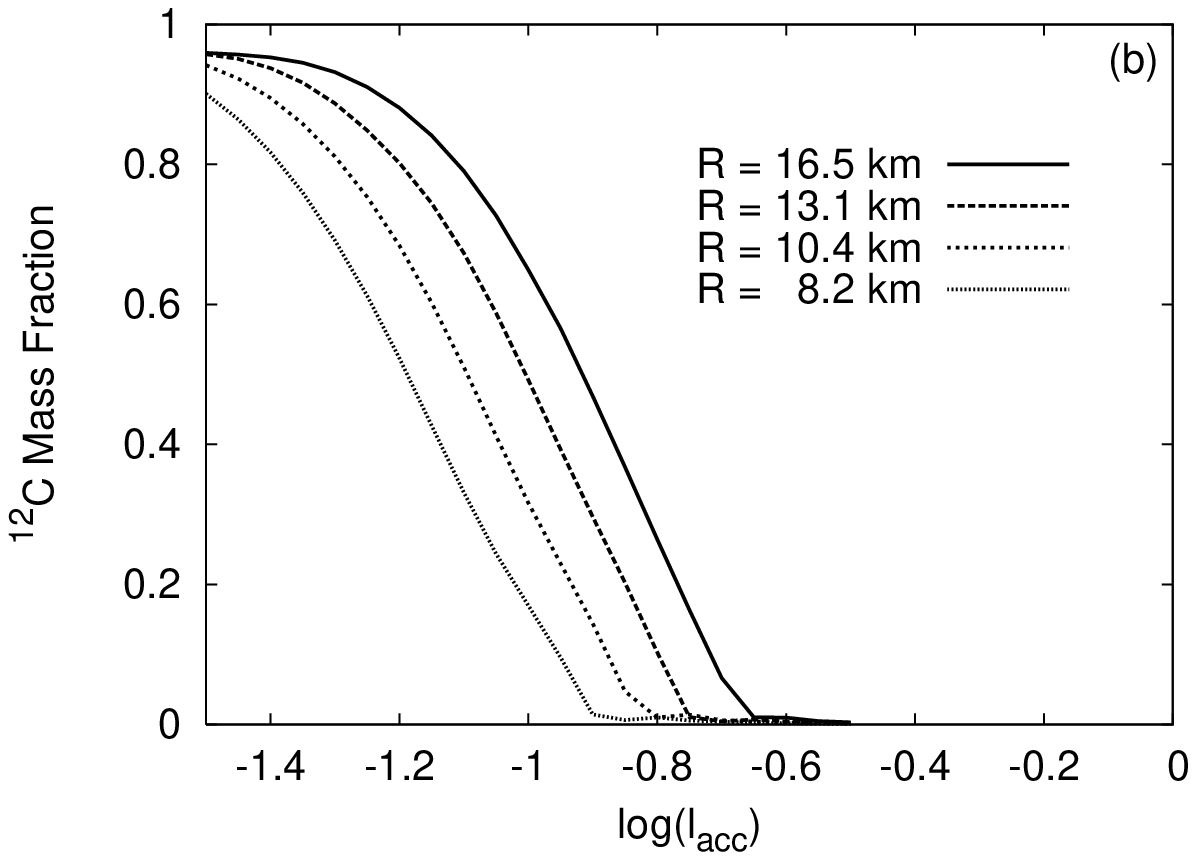}
\caption{$^{12}$C mass fraction at the the base of the accreted column
due to stable burning as a function of accretion rate.  The parameter
$l_{\mathrm{acc}}$ is the accretion rate normalized to the Eddington
limit.  Figure (a) shows the $^{12}$C mass fractions deep within the 
accreted layer of $10.4$ km
neutron stars with $Z_{\mathrm{CNO, out}} = Z_{\mathrm{CNO, \odot}}$
for each of the nine thermal structure models.  Figure (b) shows the
$^{12}$C mass fractions of neutron stars with $Z_{\mathrm{CNO, out}} =
Z_{\mathrm{CNO, \odot}}$, $Q = 5.2$ below the accreted layer, and with
cores that emit neutrinos via mUrca reactions for four different
radii.
}
\label{Cmassfrac}
\end{figure*}

For the relatively high accretion rates at which superbursts occur,
the envelope and ocean of the neutron star are thermally insulated
from the inner crust and core, and so the thermal profile of the ocean
depends primarily on the accretion rate \citep{CN05}.  Consequently,
the carbon yield resulting from stable hydrogen and helium burning is
rather insensitive to both the conductive opacity of the crust and
neutrino emission mechanism of the core, but it is quite sensitive to
the accretion rate.  We plot the $^{12}$C mass fraction as a function
of accretion rate for nine neutron stars with different conductive
opacities and core neutrino emissivities in Figure \ref{Cmassfrac}a.
The parameter $l_{\mathrm{acc}}= \dot{M} / \dot{M}_{\mathrm{Edd}}$ is
the accretion rate normalized to the Eddington limit, where
$\dot{M}_{\mathrm{Edd}} = 4 \pi G M (1+z)/c z \kappa_{\mathrm{es}}$,
$z = (1-2GM/Rc^{2})^{-1/2}-1$ is the gravitational redshift, and
$\kappa_{\mathrm{es}} = 0.4$ $\mathrm{cm}^{2}\,\mathrm{g}^{-1}$.  At
lower accretion rates, most of the hydrogen burns before the helium
ignites, so the carbon produced via stable helium burning survives.
As the accretion rate increases, the hydrogen and helium ignition
regions overlap, so the carbon produced via stable helium burning will
be processed into oxygen during the hot CNO cycle, and the oxygen will
be removed from the hot CNO cycle via breakout reactions (see
\S\ref{stableburn}).  Therefore, the carbon yield substantially
decreases with increasing accretion rate.

Unlike the conductive opacity and core neutrino emissivity, the
stellar radius significantly affects the carbon yield for a given
$l_{\mathrm{acc}}$, as illustrated in Figure \ref{Cmassfrac}b.  There
are two reasons for this.  First, the accretion rate per unit area
$\dot{\Sigma}$ is lower for a larger radius, so $\Sigma_{\mathrm{H}}$,
the column depth at which hydrogen burns, decreases with increasing
radius.  Second, the density and temperature at a given column depth
$\Sigma$ are generally lower for a larger radius, so the column depth
at which helium burns increases with increasing radius.  The
combination of these two effects implies that helium is more likely to
burn in a hydrogen-deficient medium if the stellar radius is large.
Therefore, stable burning on neutron stars with larger radii will
yield more carbon fuel.

\begin{figure*}
\plottwo{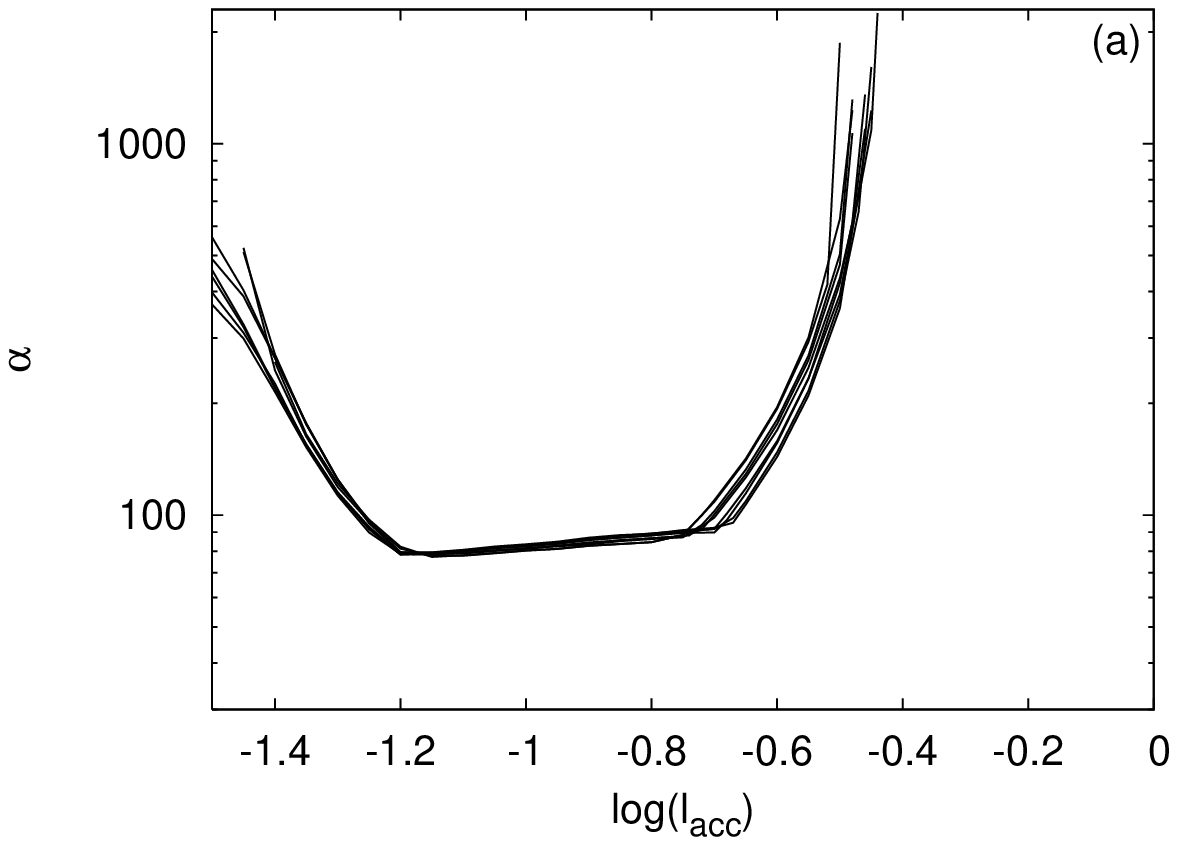}{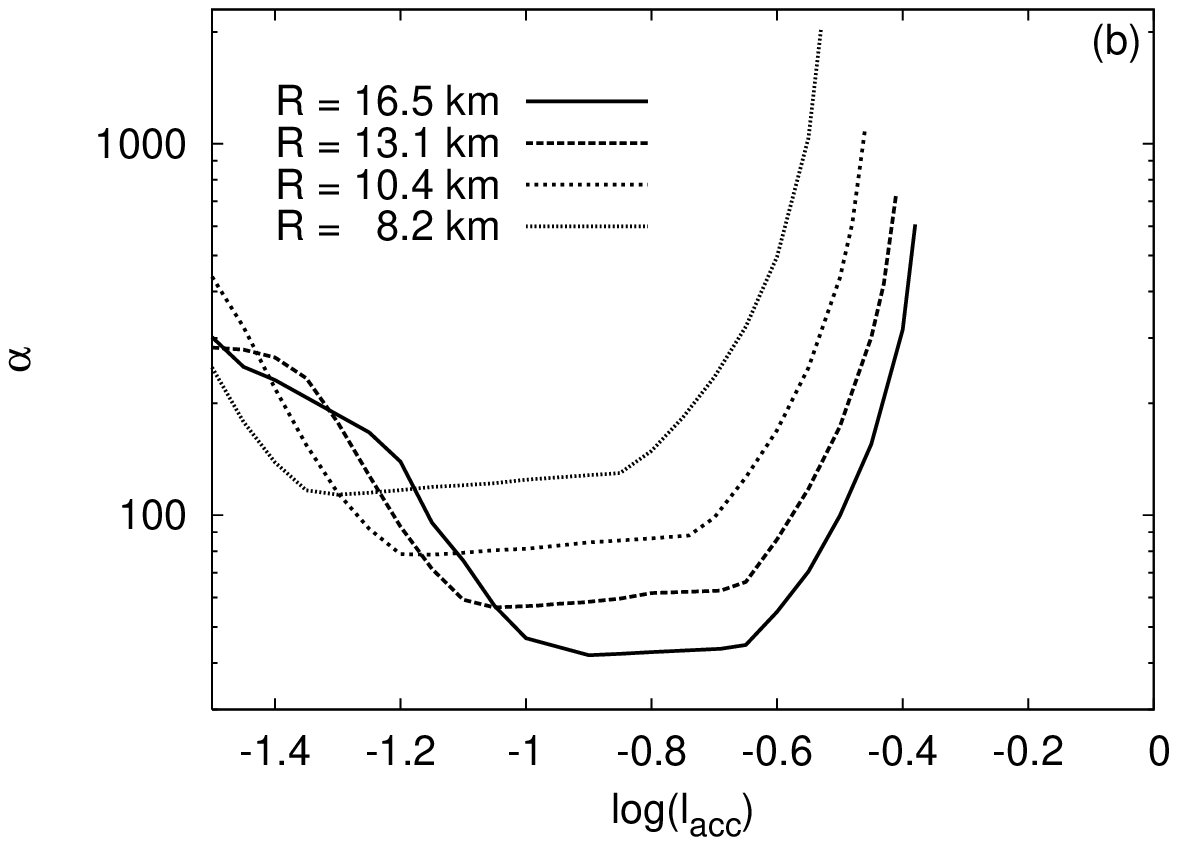}
\caption{$\alpha$-values of normal bursts as a function of accretion
rate.  Figure (a) shows $\alpha$-values of $10.4$ km neutron stars
with $Z_{\mathrm{CNO, out}} = Z_{\mathrm{CNO, \odot}}$ for each of the
nine thermal structure models.  Figure (b) shows $\alpha$-values of
neutron stars with $Z_{\mathrm{CNO, out}} = Z_{\mathrm{CNO, \odot}}$,
$Q = 5.2$ below the accreted layer, and with cores that emit neutrinos
via mUrca reactions for four different radii.
}
\label{alpha}
\end{figure*}

The derived carbon yield for a given $l_{\mathrm{acc}}$ shown in
Figure \ref{Cmassfrac} is meaningless, however, if prompt normal bursts
occur at that $l_{\mathrm{acc}}$.  We plot the $\alpha$-values of
normal bursts as a function of accretion rate for nine neutron stars 
with different conductive
opacities and core neutrino emissivities in Figure \ref{alpha}a and
with four different stellar areal radii in Figure \ref{alpha}b.  To
perform these normal burst calculations, we use the model described by
\citet{RLCN05} tailored for persistent accretors, which is a slightly
modified version of the model of \citet{CN05}.  Like the carbon yield
due to stable burning, $\alpha$ depends weakly on the thermal
structure of the crust and core, but it is quite sensitive to the
stellar radius.  For the $R = 10.4$ km models shown in Figure \ref{alpha}a, 
no normal bursts occur for $l_{\mathrm{acc}} \gtrsim 0.3$.  Delayed 
mixed bursts occur for $0.2 \lesssim l_{\mathrm{acc}} \lesssim 0.3$, 
prompt bursts occur for $0.05 \lesssim l_{\mathrm{acc}} \lesssim 0.2$, 
and delayed helium bursts occur for $l_{\mathrm{acc}} \lesssim 0.05$ 
\citep[for details on the various bursting regimes, see][]{NH03}.

We list the range of accretion rates in which superbursts occur for
each model in Table 1.  No superbursts occur at any accretion rate for
any of the 36 models for which $Z_{\mathrm{CNO, out}} =
Z_{\mathrm{CNO, \odot}}$, where $Z_{\mathrm{CNO, \odot}} = 0.016$.
For the high accretion rates at which either $\alpha \gg 100$ or
normal bursts do not occur, stable burning produces a negligible
amount of carbon, in agreement with \citet{SBCW99,SBCO03}.  For
$l_{\mathrm{acc}} \sim 0.1$, prompt bursts destroy any carbon produced
via stable burning \citep{Setal01,SBCO03,KHK04,Wetal04,FBLTW05}.  At
still lower $l_{\mathrm{acc}}$, the carbon may survive the delayed
bursts, but the carbon that remains will burn stably \citep{CB01,
CN05,CMintZP05} and no superburst will occur.  Increasing the radius
does not raise the probability that a superburst will occur.  Though
increasing the radius will increase the carbon yield due to stable
burning, it will also increase the limiting accretion rate above which
delayed mixed bursts occur.  These results hold for any
$Z_{\mathrm{CNO, out}} < Z_{\mathrm{CNO, \odot}}$ as well, for
decreasing $Z_{\mathrm{CNO, out}}$ will both lower the carbon yield
and extend the range of accretion rates over which prompt normal
bursts occur.

\begin{deluxetable*}{ccc|cccc}
\tablecolumns{7} 
\tablewidth{0pt}
\tablecaption{Results}
\tablehead{
\colhead{} & 
\colhead{} &
\colhead{} &
\multicolumn{4}{c}{$\dot{M}/\dot{M}_{\mathrm{Edd}}$ Range for Superbursts for $Z_{\mathrm{CNO, out}} =$}\\
\cline{4-7}\\
\colhead{$R$} & 
\colhead{Core $\nu$} &
\colhead{$Q$} &
\colhead{$Z_{\mathrm{CNO, \odot}}$} &
\colhead{$2 Z_{\mathrm{CNO, \odot}}$} &
\colhead{$3 Z_{\mathrm{CNO, \odot}}$} &
\colhead{$4 Z_{\mathrm{CNO, \odot}}$\tablenotemark{a}}\\
\colhead{(km)} & 
\colhead{Emissivity} &
\colhead{} &
\colhead{} &
\colhead{} &
\colhead{} &
\colhead{}
}
\startdata
$8.2$ &hot mUrca&$5.2$     &&&&\\
$8.2$ &hot mUrca&$100$     &&&$0.04$-$0.06$&$0.04$-$0.08$\\
$8.2$ &hot mUrca&disordered&&&$0.03$-$0.07$&$0.03$-$0.10$\\
$8.2$ &mUrca    &$5.2$     &&&$0.05$-$0.07$&$0.03$-$0.09$\\
$8.2$ &mUrca    &$100$     &&&$0.03$-$0.07$&$0.03$-$0.11$\\
$8.2$ &mUrca    &disordered&&&$0.03$-$0.07$&$0.03$-$0.13$\\
$8.2$ &Pion     &$5.2$     &&&$0.03$-$0.07$&$0.03$-$0.06$\\
$8.2$ &Pion     &$100$     &&&$0.03$-$0.07$&$0.03$-$0.11$\\
$8.2$ &Pion     &disordered&&&$0.03$-$0.07$&$0.03$-$0.14$\\
$10.4$&hot mUrca&$5.2$     &&$0.04$-$0.05$&$0.06$-$0.09$&$0.07$-$0.13$\\
$10.4$&hot mUrca&$100$     &&$0.03$-$0.05$&$0.06$-$0.09$&$0.07$-$0.13$\\
$10.4$&hot mUrca&disordered&&$0.03$-$0.05$&$0.06$-$0.09$&$0.07$-$0.13$\\
$10.4$&mUrca    &$5.2$     &&$0.03$-$0.05$&$0.06$-$0.09$&$0.07$-$0.13$\\
$10.4$&mUrca    &$100$     &&$0.03$-$0.05$&$0.06$-$0.09$&$0.07$-$0.13$\\
$10.4$&mUrca    &disordered&&$0.03$-$0.05$&$0.06$-$0.09$&$0.07$-$0.13$\\
$10.4$&Pion     &$5.2$     &&$0.03$-$0.05$&$0.06$-$0.09$&$0.07$-$0.13$\\
$10.4$&Pion     &$100$     &&$0.03$-$0.05$&$0.06$-$0.09$&$0.07$-$0.13$\\
$10.4$&Pion     &disordered&&$0.03$-$0.05$&$0.06$-$0.09$&$0.07$-$0.13$\\
$13.1$&hot mUrca&$5.2$     &&&&$0.10$-$0.18$, $0.22$\\
$13.1$&hot mUrca&$100$     &&&&$0.10$-$0.18$, $0.22$\\
$13.1$&hot mUrca&disordered&&&&$0.10$-$0.18$, $0.22$\\
$13.1$&mUrca    &$5.2$     &&&&$0.10$-$0.18$, $0.22$-$0.25$\\
$13.1$&mUrca    &$100$     &&&&$0.10$-$0.18$, $0.22$\\
$13.1$&mUrca    &disordered&&&&$0.10$-$0.18$, $0.22$\\
$13.1$&Pion     &$5.2$     &&&&$0.10$-$0.18$, $0.22$-$0.25$\\
$13.1$&Pion     &$100$     &&&&$0.10$-$0.18$, $0.22$-$0.25$\\
$13.1$&Pion     &disordered&&&&$0.10$-$0.18$, $0.22$-$0.25$\\
$16.5$&hot mUrca&$5.2$     &&&&$0.04$-$0.07$, $0.14$-$0.32$\\
$16.5$&hot mUrca&$100$     &&&&$0.04$-$0.07$, $0.14$-$0.32$\\
$16.5$&hot mUrca&disordered&&&&$0.04$-$0.07$, $0.14$-$0.32$\\
$16.5$&mUrca    &$5.2$     &&&&$0.04$-$0.07$, $0.14$-$0.32$\\
$16.5$&mUrca    &$100$     &&&&$0.04$-$0.07$, $0.14$-$0.32$\\
$16.5$&mUrca    &disordered&&&&$0.04$-$0.07$, $0.14$-$0.32$\\
$16.5$&Pion     &$5.2$     &&&&$0.04$-$0.07$, $0.14$-$0.32$\\
$16.5$&Pion     &$100$     &&&&$0.04$-$0.07$, $0.14$-$0.32$\\
$16.5$&Pion     &disordered&&&&$0.04$-$0.07$, $0.14$-$0.32$\\
\enddata
\tablenotetext{a}{For some models, superbursts occur over two distinct 
ranges of $\dot{M}/\dot{M}_{\mathrm{Edd}}$ separated by a stable 
zone.  Both $\dot{M}/\dot{M}_{\mathrm{Edd}}$ ranges are given in 
these cases.}

\end{deluxetable*}

\subsection{Results for $Z_{\mathrm{CNO, out}} > Z_{\mathrm{CNO, \odot}}$}

\begin{figure}
\plotone{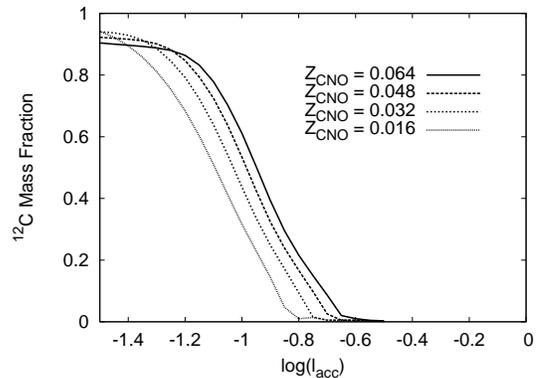}
\caption{$^{12}$C mass fraction at the the base of the accreted column
due to stable burning as a function of accretion rate for four
different values of $Z_{\mathrm{CNO, out}}$, the mass fraction of CNO
elements in the accreted matter.  $R = 10.4$ km, $Q = 5.2$ below the
accreted layer, and the stellar core is assumed to emit neutrinos via
mUrca reactions.
}
\label{CmassfracCNO}
\end{figure}

As mentioned in \S\ref{stableburn}, carbon will survive to ultimately
trigger a superburst only if the stable helium burning that produces
it takes place in a hydrogen-deficient environment.  Thus, the column
depth at which the accreted hydrogen burns, $\Sigma_{\mathrm{H}}$,
must be less than the column depth at which helium burns to provide
enough fuel for a superburst.  According to equation (\ref{sigmaH}),
$\Sigma_{\mathrm{H}}$ is inversely proportional to $Z_{\mathrm{CNO,
out}}$, so if the accreted matter is rich in CNO elements, hydrogen is
more likely to burn out before helium ignites.  Therefore, more of the
carbon produced via stable helium burning should survive.  Figure
\ref{CmassfracCNO} illustrates this effect of the CNO mass fraction of
the accreted matter on the carbon yield.  We emphasize that the
individual mass fractions of the accreted carbon, nitrogen, and oxygen
are unimportant.  Since the accreted CNO ions are simply catalysts for
hydrogen burning, it is only the sum of the three individual mass
fractions that matters. 

\begin{figure}
\plotone{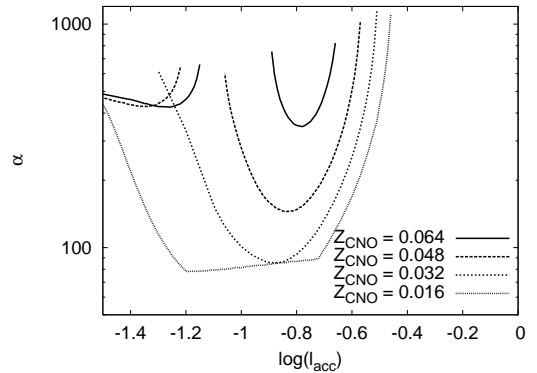}
\caption{$\alpha$-values of normal bursts as a function of
accretion rate for four different values of $Z_{\mathrm{CNO, out}}$.  
$R = 10.4$ km, $Q = 5.2$ below the accreted layer, and the stellar 
core is assumed to emit neutrinos via mUrca reactions.
}
\label{alphaCNO}
\end{figure}

As in \S\ref{resultssolar}, the derived carbon yield for a given
$l_{\mathrm{acc}}$ is meaningless if prompt normal bursts occur at
that $l_{\mathrm{acc}}$.  We plot the $\alpha$-values of normal bursts
for four neutron stars with different CNO abundances of the accreted
matter in Figure \ref{alphaCNO}.  Evidently, increasing
$Z_{\mathrm{CNO, out}}$ affects normal bursts in several ways.  First,
it lowers the critical accretion rate above which normal bursts do not
occur.  Hydrogen burning is always $\beta$-limited at the accretion
rates considered in this article, so only unstable helium burning
triggers normal bursts.  The helium nuclear energy generation rate $
\epsilon_{\mathrm{He}}$ monotonically increases with temperature, but
the temperature sensitivity decreases as temperature increases.  In
other words, $\partial^{2} \epsilon_{\mathrm{He}} / \partial T^{2}
< 0$ \citep{FL87b}.  Therefore, helium burns stably at sufficiently
high temperatures \citep{B98}.  Since the temperature of the ocean
increases with accretion rate, there exists a critical accretion rate
above which normal bursts do not occur.  The increased hydrogen
nuclear energy generation rate due to the CNO enhancement raises the
temperature of the ocean, so helium is more likely to burn in a stable
fashion, resulting in a lower critical accretion rate.  Second, the
$\alpha$-values of normal bursts are generally higher.  Just below the
critical accretion rate, the normal bursts that occur are ``mixed''
bursts, meaning that the bursts consume substantial amounts of both
hydrogen and helium.  Since $\Sigma_{\mathrm{H}}$ is lower, a larger
fraction of the hydrogen burns before helium ignites unstably, which
means that less hydrogen is consumed during the burst.  This results
in a higher $\alpha$.  Third, the region of delayed helium bursts is
curtailed, and a range of accretion rates exists below this region in
which normal bursts do not occur.  Again, this is due to the increased
temperature of the ocean due to the intensified hydrogen energy
generation rate.

We list the ranges of accretion rates at which superbursts are
triggered for $Z_{\mathrm{CNO, out}} = 2 Z_{\mathrm{CNO, \odot}}$, $3
Z_{\mathrm{CNO, \odot}}$, and $4 Z_{\mathrm{CNO, \odot}}$ in Table 1.
We find that an enhancement of the CNO abundance in the accreted
matter will allow for superbursts to occur in certain ranges of
accretion rates.  The quantity $Z_{\mathrm{CNO, out}}$ is unique in
that it is the only free parameter that will both increase the carbon
yield due to stable burning at a given accretion rate and decrease the
range of accretion rates at which prompt normal bursts occur.  Therefore, 
we suggest that an enhancement of the CNO abundance in the accreted
matter is a prerequisite for superbursts to occur in systems that 
accrete predominantly hydrogen.  

\section{Comparison with Previous Theoretical Studies}\label{comparewth}

As noted in \S\ref{resultssolar}, our conclusion that neutron stars
that accrete material with $Z_{\mathrm{CNO, out}} \lesssim
Z_{\mathrm{CNO, \odot}}$ will not exhibit superbursts in consistent
with the results of previous theoretical investigations.  However, a
comparison between Figure \ref{Cmassfrac}a of this work and Figure 2
of \citet{SBCO03} illustrates that a discrepancy exists between the
mass fraction of carbon that survives stable burning as a function of
accretion rate that we calculate and the same quantity \citet{SBCO03}
derive.  While both groups find that the carbon mass fraction $\approx
0.3$ at $l_{\mathrm{acc}} = 0.1$ (the lowest accretion rate they
consider), \citet{SBCO03} find that the mass fraction $\approx 0.08$
at $l_{\mathrm{acc}} = 0.3$, while our calculations imply that a
negligible amount of carbon will survive at this accretion rate.
Generally speaking, it appears that the resulting carbon mass fraction
is more sensitive to the accretion rate in our model than it is in the
model of \citet{SBCO03}.  In particular, Figure \ref{HHeCburning}b
illustrates that the hydrogen mass fraction plateaus around $\Sigma
\approx 10^{8}$ $\mathrm{g\,cm}^{-2}$ in our model, whereas no such
plateau exists in the model of \citet{SBCO03}.  In this section we
discuss our efforts to resolve this discrepancy.

One difference between the two models is that ours is
general-relativistic, while theirs is Newtonian.  Consequently, the
gravitational acceleration in our model is greater.  Although this
will reduce the carbon yield somewhat, the effect is too small to
account for such a large disparity.  Furthermore, we would expect
that the larger gravitational acceleration would reduce the carbon
yield at all accretion rates, which means that the results of the two
models should disagree at all accretion rates.  Thus, we deduce that
the inclusion of general relativity in our model is not the source of
the discrepancy.

Another difference between the two models is the treatment of the
inner boundary condition.  While we set the inner boundary condition
at the crust-core interface, \citet{SBCO03} set it at the base of the
burning layer.  Therefore, we have conducted experimental calculations
in which we set our boundary condition near the base of the burning
layer.  We are unable to reproduce their results at $l_{\mathrm{acc}}
= 0.3$ for any reasonable choice of boundary condition, and so we
conclude that the treatment of the inner boundary condition is not the
sole cause of the discrepancy.

A potential deficiency in our model may be that we do not directly
couple our reaction network into our energy conservation equation.
This could result in an inaccurate thermal profile which would affect
our carbon yield.  As discussed in \S\ref{Cabundance}, we have carried
out calculations, specifically at $l_{\mathrm{acc}} = 0.3$, in which
the nuclear energy generation rate of our reaction network is directly
coupled into equation (\ref{energyequation}), and the final carbon
yield differs negligibly from our old result.  Different radiative 
opacity prescriptions would affect the thermal profile too, but we 
use the same prescription as \citet{SBCO03}, so the opacity is likely 
not an issue either.

\begin{figure}
\plotone{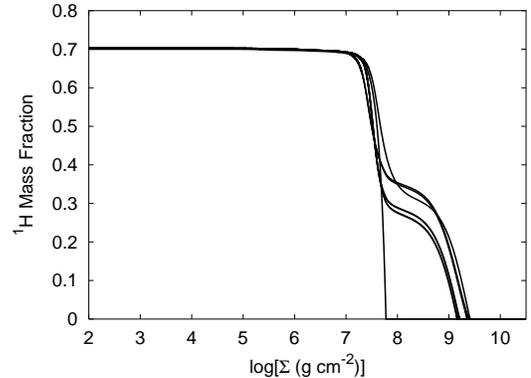}
\caption{$^{1}$H mass fraction profiles in the accreted column due to
stable burning at $\dot{M} = 0.3 \dot{M}_{\mathrm{Edd}}$ for all
possible combinations of the most recent
$^{15}$O($\alpha$,$\gamma$)$^{19}$Ne and $^{14}$O($\alpha$, p)$^{17}$F
reaction rates.  The bottommost profile, where $^{1}$H is depleted at
$\Sigma \approx 10^{7.8}$ $\mathrm{g\,cm}^{-2}$, is the result of a
calculation in which we artificially lowered both reaction rates by a
factor of $10$ and $100$, respectively.
}
\label{Hbumps}
\end{figure}

A plausible explanation for the discrepancy which we are not able to
rule out is that some of the corresponding reactions rates in the two
networks differ significantly.  The two networks are of similar
vintage, however, so we presume that none of the corresponding
reaction rates disagrees considerably.  Unfortunately, since we do not
have access to the network of \citet{SBCO03}, we are unable to conduct
a detailed comparison.  However, two particular reactions that would
certainly affect the final carbon yield and the hydrogen mass fraction
profile are the CNO breakout reactions $^{15}$O($\alpha$,
$\gamma$)$^{19}$Ne and $^{14}$O($\alpha$, p)$^{17}$F
\citep{SBCW99,FGWD04}.  To investigate their effects, we repeated the
calculation shown in Figure \ref{HHeCburning}b using all possible
combinations of the three most recent $^{15}$O($\alpha$,
$\gamma$)$^{19}$Ne rates and the two most recent $^{14}$O($\alpha$,
p)$^{17}$F rates (J. L. Fisker 2005, private communication).  Figure
\ref{Hbumps} shows the resulting hydrogen mass fraction profiles.  The
CNO breakout reactions extract $^{14}$O and $^{15}$O from the CNO
cycle, causing the hydrogen burning to stall.  Hydrogen thus survives
to a greater depth, which reduces the carbon yield.  Although the
lower $^{15}$O($\alpha$, $\gamma$)$^{19}$Ne and $^{14}$O($\alpha$,
p)$^{17}$F reaction rates curtail the plateau somewhat, we find that
none of the six possible reaction rate combinations eliminates the
plateau entirely.  Furthermore, all of the final carbon yields differ
by less than a factor of $2$.  We are able to remove the hydrogen
plateau only by artificially reducing the $^{15}$O($\alpha$,
$\gamma$)$^{19}$Ne and $^{14}$O($\alpha$, p)$^{17}$F reaction rates by
a factor of $10$ and $100$, respectively.  When we do this, we find
that the carbon mass fraction $\approx 0.08$ for $l_{\mathrm{acc}} =
0.3$.

In summary, we are unable to pinpoint the reason why the final carbon
yield due to stable burning at relatively high accretion rates derived
from our model differs from the yield derived by \citet{SBCO03}.
Fortunately, our general conclusions remain unchanged regardless of
which model is a better description of the nuclear reactions that
occur in nature, for even if the model of \citet{SBCO03} is
``correct,'' we still find that the mass donors in systems that
exhibit superbursts must be evolved stars with enhanced CNO abundances
in order to both generate sufficient amounts of carbon fuel and
reproduce the observed characteristics of the normal bursts.

\section{Comparison with Observations}\label{comparewobs}

The nine superbursts that have been observed as of this writing,
excluding those from the anomalous system GX 17+2, occurred in systems
with accretion rates between $10\%$ and $25\%$ of the Eddington limit.
The normal bursts that these systems exhibit differ in two ways from
the normal bursts in other systems with similar accretion rates.
First, the normal bursts are delayed bursts, with $\alpha \gtrsim
1000$ \citep{KHvdKLM02,intZetal03,intZetalHEAD04}, whereas most
systems accreting at these rates exhibit prompt bursts, with $\alpha <
100$ \citep{vPPL88,NH03}.  Second, the average durations of normal
bursts that occur in these systems are shorter than those of normal
bursts in systems that do not exhibit superbursts
\citep{intZetal03,intZetalHEAD04}.  Any successful theoretical model
of superbursts must be able to explain these facts.  In this
investigation, we concentrate on systems in which the accreted matter
is predominantly hydrogen.  This excludes the system 4U 1820-303, in
which the accreted matter is probably dominated by helium
\citep{SB02,C03}.  It is likely that the long periods of stable helium
burning that take place in the hydrogen-deficient ocean of this source
produce large amounts of carbon fuel for superbursts.  We note,
however, that it is not understood theoretically why this system does
not exhibit normal bursts during these periods, when the accretion
rate is near its maximum \citep{B95,B98,C03}.

We begin by discussing the range of accretion rates in which
superbursts have been observed, $l_{\mathrm{acc}} \approx 0.1$-$0.25$.
From Table 1, we require that $Z_{\mathrm{CNO, out}} \gtrsim 4
Z_{\mathrm{CNO, \odot}}$ in order for a superburst to occur in this
range.  $Z_{\mathrm{CNO, out}}$ is the most important parameter that
determines whether a system in this range of accretion rates will
exhibit a superburst.  If the CNO mass fraction of the accreted matter
is low, we find that there is no scenario in which a sufficient amount
of carbon fuel will survive to ultimately trigger a superburst.  The
second most important parameter is the stellar areal radius.  Although
a neutron star with almost any plausible radius can exhibit a
superburst at $l_{\mathrm{acc}} \approx 0.1$ given a sufficiently
large value of $Z_{\mathrm{CNO, out}}$, neutron stars will not exhibit
superbursts at $l_{\mathrm{acc}} \approx0.2$ unless the radius $R
\gtrsim 13$ km.  If all neutron stars are indeed rather large, then
our model predicts that superbursts will not occur in systems that
accrete predominantly hydrogen if $l_{\mathrm{acc}} < 0.1$, in
agreement with observations.  Our model may also explain why
superbursts are not observed in most low mass X-ray binaries with
$l_{\mathrm{acc}} \gtrsim 0.3$ (excluding the anomalous GX 17+2, which
we will discuss below).

Previous theoretical models of normal bursts all produce prompt bursts
at $l_{\mathrm{acc}} \approx 0.1$, with $\alpha \lesssim 100$.  The
results of these models are inconsistent with the delayed bursts
observed at $l_{\mathrm{acc}} \approx 0.1$ from the systems 4U
1636-536, KS 1731-260, and 4U 1254-690, in which $\alpha \approx 440$,
$780$, and $4800$, respectively \citep{intZetal03,intZetalHEAD04}.
These models usually assume that $Z_{\mathrm{CNO, out}} \lesssim
Z_{\mathrm{CNO, \odot}}$.  However, \citet{TM92} showed that
theoretical models of normal bursts from the superburster 4U 1636-536
are inconsistent with observations unless a significant amount of CNO
exists in the ocean.  Figure \ref{alphaCNO} shows that an enhancement
of the CNO abundance in the accreted matter produces delayed bursts at
$l_{\mathrm{acc}} \approx 0.1$ that are consistent with observations.
The systems Ser X-1, GX 3+1, and 4U 1735-444 have accretion rates
$l_{\mathrm{acc}} \approx 0.2$, $0.2$, and $0.25$ \citep{K04}, and
they exhibit normal bursts with $\alpha \approx 5800$, $2100$, and
$4400$, respectively \citep{intZetal03,intZetalHEAD04}.  Figure
\ref{alphaCNO} shows that delayed bursts occur at these accretion
rates only if the CNO abundance in the accreted matter is high.
Therefore, our theoretical models of normal bursts are consistent with
the observed delayed bursts with $\alpha \gtrsim 1000$ only if the CNO
abundance in the accreted matter in notably greater than solar.

\begin{figure}
\plotone{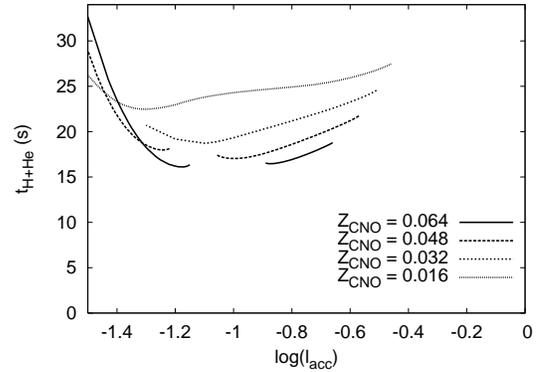}
\caption{Effective normal burst duration as a function of accretion
rate, calculated by dividing the total burst energy by the Eddington
luminosity, for four different values of $Z_{\mathrm{CNO, out}}$.  $R
= 10.4$ km, $Q = 5.2$ below the accreted layer, and the stellar core
is assumed to emit neutrinos via mUrca reactions.  Note that
$t_{\mathrm{H+He}}$ decreases with increasing $Z_{\mathrm{CNO, out}}$ 
for $l_{\mathrm{acc}} \gtrsim 0.06$.
}
\label{tHHeCNO}
\end{figure}

The average decay times of normal bursts in all systems that exhibit
superbursts are lower than the average decay times of normal bursts in
systems that do not exhibit superbursts
\citep{intZetal03,intZetalHEAD04}.  We plot the effective normal burst
duration $t_{\mathrm{H+He}}$ as a function of $l_{\mathrm{acc}}$ in
Figure \ref{tHHeCNO}, where $t_{\mathrm{H+He}}$ equals the energy
released via hydrogen and helium burning during a normal burst divided
by the Eddington luminosity \citep[see][]{NH03}.  Figure \ref{tHHeCNO}
illustrates that systems with a large CNO abundance in the accreted
matter will exhibit normal bursts with smaller durations.  Again, our
models of normal bursts are consistent with observations only if the
CNO abundance in the accreted matter in high.  

Several of the systems have $\alpha$-values that are even higher than
those plotted in Figure \ref{alphaCNO}.  This could easily occur if
the systems have accretion rates close to a critical accretion rate
that separates stable burning and unstable burning \citep{vPCLJ79}.
The systems 4U 1636-536, KS 1731-260, Ser X-1, GX 3+1, and 4U 1735-444
all undergo episodes of irregular bursting behavior, which implies
that they are accreting near such a critical rate
\citep{Letal87,MFMB02,SBCML83,dHetal03,vPPLST88}.  Previous
theoretical models predict that normal bursts will occur for all
accretion rates below the critical rate above which helium burning is
stable.  \citet{NH03} find that the critical accretion rate is roughly
$30\%$ of the Eddington limit, which is consistent with observations
\citep{Cetal03,RLCN05}.  Thus it is not surprising that Ser X-1, GX 3+1, and
4U 1735-444 experience this irregular behavior, for their accretion
rates are close to this upper bound.  What is surprising is that 4U
1636-536 and KS 1731-260 experience irregular bursting behavior too,
since $l_{\mathrm{acc}} \approx 0.1$ for these two systems.  This
implies that the range of accretion rates at which normal bursts occur
is not continuous, for there exists a range of accretion rates below
$l_{\mathrm{acc}} \approx 0.1$ in which normal bursts do not occur.
The bursting behavior of KS 1731-260, which is a transient system, is
particularly interesting.  \citet{MFMB02} find that short,
photospheric radius expansion bursts occur at high accretion rates, where
$l_{\mathrm{acc}} \sim 0.1$, and long bursts with no evidence of
radius expansion occur at low accretion rates, where $l_{\mathrm{acc}} \sim
0.01$, but no bursts occur at intermediate accretion rates.  All of this is
nicely reproduced in Figures \ref{alphaCNO} and \ref{tHHeCNO} for
models with CNO enhancement in the accreted matter.  \citet{SBCCT05}
suggest that the accretion rate of 4U 1636-536 is currently decreasing
with time.  Our models predict that in this system in the near future
either $\alpha$ will rise or normal bursts will cease altogether.
This prediction should be testable with further observations.

The system GX 17+2, which has an accretion rate $l_{\mathrm{acc}}
\approx 0.8$ and exhibits normal bursts with $\alpha \sim 1000$ as
well as superbursts, is still a mystery to us.  Though there is little
direct evidence that the accretion rate is so high, \citet{intZCC04}
conclude with reasonable certainty that the accretion luminosity is
always greater than $60\%$ of the Eddington limit.  If this is indeed
true, then clearly the observations of GX 17+2 are inconsistent with
our results.  Theoretical models predict that systems that accrete at
this rate will show superbursts given a sufficient amount of carbon
\citep{CB01,B04,CN05}.  However, it is not understood why so much
carbon should survive or why normal bursts should occur at such a high
accretion rate.  We offer two possible explanations, neither of which
is without issues.  First, we find that a reduction of the radiative
opacities by roughly an order of magnitude will produce delayed mixed
bursts at $l_{\mathrm{acc}} \approx 0.8$ with $\alpha \sim 1000$ and
substantially increase the carbon yield due to stable burning.  GX
17+2 has a magnetic field that is much stronger than those of typical
low mass X-ray binaries \citep{KHvdKLM02}, and strong internal radial
magnetic fields lower the radiative opacities of the ocean
\citep{M92,L00}.  However, if an internal magnetic field were solely
responsible for the opacity reduction, it would have to be $\sim
10^{13}$ G \citep{vR88}, whereas \citet{Wetal96} derive an upper limit
of $\approx 5 \times 10^{9}$ G for this system.  Therefore, we think
this explanation is highly unlikely.  Second, we find that lowering
the hydrogen mass fraction of the accreted matter, $X_{\mathrm{out}}$,
by a factor of $\sim 2$ will produce delayed mixed bursts at
$l_{\mathrm{acc}} \approx 0.8$ with $\alpha \sim 1000$ and
substantially increase the carbon yield due to stable burning.  This
will lower the burst durations as well.  This explanation is
definitely plausible, for theoretical evolutionary models of
intermediate-mass X-ray binaries suggest that the secondaries will be
hydrogen-poor \citep{PRP02,PRP04}.  Though some of the normal bursts
observed in GX 17+2 were quite short ($\sim 10$ seconds), most of the
observed normal bursts were rather long ($\sim 10$ minutes), which
implies that hydrogen was abundant in the ocean when the bursts
occurred \citep{KG84,THKMM84,Setal86,Ketal97,KHvdKLM02}.  This issue
should be investigated further.

\section{Observational Evidence of CNO Enhancement}\label{CNOenhancement}

Evidence that the accreted material in compact binaries can have
non-solar abundances is found in a variety of systems.  For example,
UV and X-ray spectroscopy of accreting white dwarfs has revealed
anomalous N/C emission line ratios in the accreted material in a
number of systems \citep[e.g.][and references
therein]{Getal03,B_BM04,Retal05}.  Since the abundances of the
accreting material directly reflect the abundances of the mass donor
star, various evolutionary scenarios have been proposed to explain
these significant deviations from solar abundances. Given that
hydrogen-rich donor stars will produce CNO deep within their cores,
CNO processing as the mass donor star burns hydrogen along its main
sequence will naturally produce CNO-enriched gas, with most of the C
converted into $^{13}$N during the CNO cycle due to the long
$\beta$-decay timescale of $^{13}$N.  If these products reach the
surface, we may expect to see significant non-solar abundances in the
accretion flow. For example, \citet{Tetal02} find an expected
$(\mathrm{N}/\mathrm{N}_{\odot}) \approx 5$-$8$ if the mass donor star
in the cataclysmic binary QZ Ser had undergone significant hydrogen
burning before mass transfer onto the white dwarf had
started. Alternatively, the mass donor may have lost (part of) its
hydrogen envelope, exposing its CNO core and significantly boosting
the CNO mass fraction of the accreting matter.  Enhanced CNO
abundances have also been reported for the black-hole X-ray binary XTE
J1118 \citep{HHKS02}, and \citet{J-GRLH05} report $4 <
(\mathrm{N}/\mathrm{O}) / (\mathrm{N}/\mathrm{O})_{\odot} < 9$ in the
high-mass X-ray binary Her X-1.  Other possible, though perhaps less
likely, avenues by which the mass donor stars can become CNO-enriched
include carbon accretion onto the donor star during the asymptotic
giant branch phase of the neutron star progenitor \citep{dKG95,SS05}
and ejecta capture from the supernova that begot the neutron star
\citep[e.g.][]{IRBCM99}.  Not many abundance analyses for X-ray
binaries have been published, but it appears that the accretion of
CNO-enriched material occurs in a number of compact binaries harboring
evolved mass donor stars.  We note that that the soft X-ray transient
4U 1608-522, in which \citet{RM05} observed a likely superburst, may
contain an evolved secondary star \citep{WHBCK02}.

Unfortunately, abundance analyses for the systems that exhibit
superbursts have not yet been reported. The large number of strong C,
N, and O spectral lines in the UV regime make it the best window for
such a study, but these analyses are severely impeded by interstellar
extinction at those wavelengths.  Strong H and He lines tend to
dominate the optical spectroscopy, and the fact that these systems
persistently accrete at $0.1 \lesssim l_{\mathrm{acc}} \lesssim 0.25$
hampers the study of the photospheric composition of the donor star
since the accretion light dominates.  For such systems, the Bowen
fluorescence blend near $\lambda$$4640$-$4660$ \AA~has proven to be a
useful indirect probe of the mass donor star \citep{SC02}.  This blend
consists of several \ion{N}{3} components that are part of the Bowen
fluorescence process, as well as nearby \ion{C}{3} lines. Irradiation
of the mass donor star leads to sharp and resolvable emission
components throughout this blend originating from its surface. This
was first demonstrated in Sco X-1, where the N and C components from
the donor were detected at similar strengths \citep[see Figure 1
in][]{SC02}.  \citet{Hetal04} report optical spectroscopy of the
superburst source Ser X-1 and remark that only the \ion{N}{3} lines
were detected in the Bowen emission blend, and no \ion{C}{3} emission
was evident.  \citet{Cetal04} present a compilation of Bowen blend
profiles from a number of X-ray binaries including the superburst
sources V801 Ara (also known as 4U 1636-536) and V926 Sco (also known
as 4U 1735-444). Again, in V801 Ara sharp components are detected from
the \ion{N}{3} components, but no sharp \ion{C}{3} components are
evident. For V926 Sco, a spectral feature coinciding with one
\ion{C}{3} line is detected, but the second \ion{C}{3} line is not
detected making it less clear-cut for this source.

We remark that, since the \ion{N}{3} transitions are part of the Bowen
fluorescence process, whereas the \ion{C}{3} are not, this blend
cannot be used for quantitative abundance analysis. However, the
dominance of N over C in two of the three observed superburst systems
is certainly suggestive that the accreting matter may indeed be
CNO-enriched.  A more detailed abundance analysis of these systems
could test whether the composition of the accreting matter does in fact
conform to the expectations of our model calculations.

\section{Summary and Conclusions}\label{summary}

By merging the theoretical superburst model of \citet{CN05} with the
detailed nuclear reaction network of \citet{MC00}, we have
investigated the physical scenarios in which accreting neutron stars
can both produce and preserve sufficient amounts of carbon fuel to
trigger superbursts.  We have constructed a total of 144 different
models that span the possible ranges of neutron star thermal
conductivities, core neutrino emission mechanisms, and areal radii, as
well as the CNO abundances in the accreted material.  We find that
neutron stars that accrete hydrogen-rich material with CNO mass
fractions less than or equal to that of the Sun will not exhibit
superbursts, regardless of their accretion rates, conductivities, core
neutrino emissivities, or radii.  Neutron stars that accrete material
with CNO mass fractions $\gtrsim 4 Z_{\mathrm{CNO, \odot}}$ can
exhibit superbursts at accretion rates in the observed range, but only
if the stellar radii are sufficiently large.  We remark that the
accreted carbon, nitrogen, and oxygen serve only as catalysts for
hydrogen burning, and therefore only the sum of their individual mass
fractions is important, not the individual mass fractions themselves.

Observationally, systems that exhibit superbursts differ from systems
that do not exhibit superbursts not only in the amount of carbon that
survives deep in the oceans of the neutron stars, but also in the
nature of their normal bursts.  Comparatively, the normal bursts
observed in systems that exhibit superbursts are much shorter in
duration, and they have $\alpha$-values that are much greater.
Increasing the CNO abundance of the accreted material both decreases
the durations of normal bursts and increases their $\alpha$-values, in
agreement with observations.  This is the first study in which the
behavior of normal Type I X-ray bursts has been incorporated into
models of systems that exhibit superbursts.  Observers have discovered
many compact stellar X-ray sources in which the accreted material is
significantly non-solar.  However, abundance analyses of the accreted
material in systems that exhibit superbursts currently do not exist.
We suggest that the secondaries of the low-mass X-ray binary systems
Ser X-1, 4U 1254-690, 4U 1735-444, GX 3+1, KS 1731-260, and 4U
1636-536 are all rich in CNO elements.  Although there is some
indirect evidence in a few of these systems that the mass donor stars
have indeed undergone CNO-processing, more observations are required
to either verify or refute this assertion. 

One issue we have not addressed is the disparity between the
superburst energies and recurrence times that have been observed and
the energies and recurrence times derived from theoretical models.
Specifically, all models predict that the energies and recurrence
times should be roughly an order of magnitude larger than those
observed.  Theory and observations still disagree even if the thermal
conductivity of the crust is low and the core temperature is high.
\citet{B04} notes that observations of KS 1731-260 imply that the
neutron star in this system in fact has a high thermal conductivity
and a low core temperature \citep{WGvdKM02}, which illustrates the
severity of this discrepancy.  Furthermore, \citet{CMintZP05} show that
neutrino emission via Cooper pairing of superfluid neutrons in the
inner crust limits the temperature of the ocean to values well below
the $\approx 6 \times 10^{8}$ K needed to trigger superbursts that
match observations, and they suggest that an additional heating
mechanism is required.  Our models do not provide for such a
mechanism, for we find that any additional heating due to the
increased CNO abundance of the accreted material has a negligible
effect on superburst energies and recurrence times.  Superbursts may
potentially be excellent probes into the interiors of neutron stars.
Further progress in the physics of superbursts, combined with better
observations, may ultimately lead to better understanding of not only
neutron star interiors, but also fundamental physics.

\acknowledgments
We would like to thank Jorge Casares, Paul Green, Josh Grindlay,
Hendrik Schatz, Jeffery McClintock, and Charles Steinhardt for helpful
discussions, Jacob Fisker for kindly providing updated reaction rates,
and the referee for useful comments and suggestions.
D.~S. acknowledges a Smithsonian Astrophysical Observatory Clay
Fellowship.  This work was supported in part by grant NNG04GL38G from
NASA.


\clearpage


\end{document}